\documentclass[nofootinbib,
reprint,prd,nopacs,
nobibnotes,
amsmath,amssymb,
aps,
]{revtex4-1}

\usepackage{graphicx}
\usepackage{dcolumn}
\usepackage{bm}

\usepackage{amsmath}
\usepackage{graphicx}
\usepackage{amsfonts}
\usepackage{latexsym}
\usepackage{bbold}
\usepackage{wasysym}
\usepackage{calligra}
\usepackage{float}
\usepackage{ulem}
\usepackage{inputenc}
\usepackage{xspace}
\usepackage{url}
\usepackage{epstopdf}
\usepackage{tikz}
\usepackage{amsthm}

\usepackage{inputenc}

\newtheorem{theorem}{Theorem}

\newcommand{\be}{\begin{equation}}
\newcommand{\ee}{\end{equation}}
\newcommand{\beq} {\begin{equation}}
\newcommand{\eeq} {\end{equation}}
\newcommand{\ba}{\begin{eqnarray}}
\newcommand{\ea}{\end{eqnarray}}

\begin{document}

	\title{Cosmic Acceleration with Torsion and Non-metricity in Friedmann-like Universes}
	
	\author{Damianos Iosifidis}
	\affiliation{Institute of Theoretical Physics, Department of Physics
		Aristotle University of Thessaloniki, 54124 Thessaloniki, Greece}
	\email{diosifid@auth.gr}
	
	\date{\today}
	\begin{abstract}
		
Starting from the generalized Raychaudhuri equation with torsion and non-metricity, and considering an FLRW spacetime we derive the most general form of acceleration equation in the presence of torsion and non-metricity. That is we derive the cosmic acceleration equation when the non-Riemannian degrees of freedom are also taken into account. We then discuss some conditions under which torsion and non-metricity accelerate/decelerate the expansion rate of the Universe. 
		
	\end{abstract}
	
	\maketitle
	
	\allowdisplaybreaks
	
	
	\tableofcontents
	
	\section{Introduction}
	\label{intro}
	
At the heart of the Cosmological description of FLRW Universes lies the so called cosmic acceleration equation or oftentimes referred to as the Cosmological Raychaudhuri equation. This very equation  determines the sign of the second derivative of the scale factor $\ddot{a}$ and therefore governs the acceleration/deceleration   of the Universe. However, it is well known that in the Riemannian geometry of Einstein's Gravity all standard forms of matter always slow down the expansion, that is to say for standard matter $\ddot{a}<0$ always (see for instance \cite{ellis2012relativistic,tsagas2008relativistic}). This is in conflict with observations which indicate accelerated expansion at both early and late times. Given that one stays in the realm of GR, the way to circumvent this problem is to consider exotic forms of matter for which $\rho+3p<0$, the so-called dark energy \cite{ellis2012relativistic,tsagas2008relativistic}. 

	However, it is interesting to note that the given form of the acceleration equation depends strongly on the simplified Riemannian geometry upon which GR is developed. It is then natural to ask, how does the acceleration equation modify when non-Riemannian\footnote{A non-Riemannian Geometry \cite{eisenhart2012non} is a generalization of the usual Riemannian one when the space is also endowed with torsion and non-metricity along with curvature. } degrees of freedom are also taken into account? Put another way, we pose the question: What is the most general form of the Cosmic Acceleration Equation (aka Raychaudhuri equation) with torsion and non-metricity in FLRW Universes? The purpose of this letter is to answer exactly this question. In particular we derive for the first time the most general form of the cosmic acceleration equation in non-Riemannian Geometries (i.e when apart from curvature, the torsion and non-metricity of space are also taken into account). We state and prove our result as a Theorem. Having obtained the acceleration equation for the generalized geometries we then apply our Theorem in constrained geometries such as Einstein-Cartan, Einstein-Weyl-Cartan\footnote{Recall that an Einstein-Cartan geometry is defined by a manifold having both curvature and torsion but vanishing non-metricity. On the other hand, an Einstein-Weyl-Cartan geometry allows also for non-metricity of the Weyl type along with curvature and torsion. Allowing also for arbitrary non-metricity we have the generalized non-Riemannian geometry. For a nice discussion on the various studies of non-Riemannian Cosmologies see \cite{puetzfeld2008probing}. We should note however that all of these studies were based on constrained geometries like Einstein-Cartan etc.} and also discuss the general case.

	\section{The setup}
	We shall consider a generalized $n-dim$ non-Riemannian Geometry, where the space apart from curvature possesses also torsion and non-metricity. We will use the definitions and notations of \cite{iosifidis2020cosmological} so we will go through the basic setup rather briefly here and refer the reader to \cite{iosifidis2020cosmological} for more details. We consider a Metric-Affine manifold endowed with a metric $g_{\mu\nu}$ and an affine connection $\Gamma^{\lambda}_{\;\;\;\mu\nu}$. We define the curvature, torsion and non-metricity tensors according to
	\beq
		R^{\mu}_{\;\;\;\nu\alpha\beta}:= 2\partial_{[\alpha}\Gamma^{\mu}_{\;\;\;|\nu|\beta]}+2\Gamma^{\mu}_{\;\;\;\rho[\alpha}\Gamma^{\rho}_{\;\;\;|\nu|\beta]}
	\eeq
	\beq
	S_{\mu\nu}^{\;\;\;\lambda}:=\Gamma^{\lambda}_{\;\;\;[\mu\nu]}
	\eeq
	\beq
	Q_{\alpha\mu\nu}:=- \nabla_{\alpha}g_{\mu\nu}
	\eeq
	and the deviation of the affine connection $\Gamma^{\lambda}_{\;\;\;\mu\nu}$ from the Levi-Civita one defines the distortion tensor \cite{schouten2013ricci}
	\begin{gather}
	N^{\lambda}_{\;\;\;\;\mu\nu}:=\Gamma^{\lambda}_{\;\;\;\mu\nu}-\widetilde{\Gamma}^{\lambda}_{\;\;\;\mu\nu}= \nonumber \\
	\frac{1}{2}g^{\alpha\lambda}(Q_{\mu\nu\alpha}+Q_{\nu\alpha\mu}-Q_{\alpha\mu\nu}) -g^{\alpha\lambda}(S_{\alpha\mu\nu}+S_{\alpha\nu\mu}-S_{\mu\nu\alpha}) \label{N}
	\end{gather}
where $\widetilde{\Gamma}^{\lambda}_{\;\;\;\mu\nu}$ is the usual Levi-Civita connection. Once the distortion is given, torsion and non-metricity can be easily computed through the relations (see for instance \cite{iosifidis2019metric})
	\beq
	S_{\mu\nu\alpha}=N_{\alpha[\mu\nu]}\;\;,\;\;\; Q_{\nu\alpha\mu}=2 N_{(\alpha\mu)\nu} \label{N1}
	\eeq
	Our definitions for the torsion vector and pseudo-vector are
	\beq
	S_{\mu}:=S_{\mu\lambda}^{\;\;\;\;\lambda} \;\;, \;\;\;
	\tilde{S}_{\mu}:=\epsilon_{\mu\alpha\beta\gamma}S^{\alpha\beta\gamma} 
	\eeq
	respectively. Note that the former is defined for any dimension while the latter only for $n=4$. As for non-metricity, we define the Weyl  and the second non-metricity vector according to
	\beq
	Q_{\alpha}:=Q_{\alpha\mu\nu}g^{\mu\nu}\;,\;\; \tilde{Q}_{\nu}=Q_{\alpha\mu\nu}g^{\alpha\mu}
	\eeq
	Finally,  as usual we define the Ricci tensor by
		\beq
	R_{\nu\beta}:=R^{\mu}_{\;\;\;\nu\mu\beta}	
	\eeq
	which is not symmetric in general.
	As a last note regarding the geometrical setup let us recall that in a general non-Riemannian Geometry (where the Metric-Affine Theories of Gravity are developed) each quantity can by decomposed into its Riemannian and non-Riemannian counterparts by virtue of $(\ref{N})$. For instance, let us denote with $\nabla_{\mu}$ and $\tilde{\nabla}_{\mu}$ the covariant derivative with respect to the  general affine-connection and the Levi-Civita\footnote{Quantities with\;  $\widetilde{}$\;  will always denote Riemannian parts (i.e. computed with respect to the Levi-Civita connection) unless otherwise stated.} one respectively. Define the scalar quantity \cite{iosifidis2018raychaudhuri}
	\beq
\Theta:=g^{\mu\nu}\nabla_{\mu}u_{\nu}	
	\eeq
	Then by using $(\ref{N})$ it is trivial to show that
	\beq
	\Theta:=\widetilde{\Theta}+\Big( -\tilde{Q}_{\mu}+1/2 Q_{\mu}+2S_{\mu}\Big)u^{\mu} \label{expan}
	\eeq
	where $\widetilde{\Theta}:=\widetilde{\nabla}_{\mu}u^{\mu}$. This decomposition will be of great use when proving our Theorem.

	\section{Cosmology with Torsion/Non-metricity}
	Let us consider an FLRW Cosmology, with the usual Robertson-Walker line element\footnote{Even though we consider a spatially flat Universe $(K=0)$ our results hold true even for curved Universes since it is well known that the curvature does not affect the acceleration equation. That is $K$ is totally absent from the so-called  second Friedmann (acceleration) equation. }
	\beq
	ds^{2}=-dt^{2}+a^{2}(t)\delta_{ij}dx^{i}dx^{j}
	\eeq
 where 	$i,j=1,2,...,n-1$ and $a(t)$ is as usual the scale factor of the Universe. In addition let $u^{\mu}$ represent the normalized $n$-velocity field of a given fluid which in co-moving coordinates is expressed as $u^{\mu}=\delta^{\mu}_{0}=(1,0,0,...,0)$. Accordingly we define the projection tensor in the usual manner
	\beq
	h_{\mu\nu}:=g_{\mu\nu}+u_{\mu}u_{\nu}
	\eeq
	 projecting objects on the space orthogonal to $u^{\mu}$. It is known that in such a highly symmetric spacetime, torsion has only $2$ (or $1$ for $n\neq 0$)  degrees of freedom \cite{tsamparlis1979cosmological} and non-metricity contributes $3$ \cite{minkevich1998isotropic}. In \cite{iosifidis2020cosmological} it is shown that the most general covariant forms of the torsion and non-metricity tensors, in such a spacetime, are given by
		\beq
	S_{\mu\nu\alpha}^{(n)}=2u_{[\mu}h_{\nu]\alpha}\Phi(t)+\epsilon_{\mu\nu\alpha\rho}u^{\rho}P(t)\delta_{n,4} \label{Scosm}
	\eeq
		\beq
	Q_{\alpha\mu\nu}=A(t)u_{\alpha}h_{\mu\nu}+B(t) h_{\alpha(\mu}u_{\nu)}+C(t)u_{\alpha}u_{\mu}u_{\nu}  \label{Qcosm}
	\eeq
with the five functions $\Phi,P,A,B,C$ representing the behaviour of the non-Riemannian degrees of freedom. As for the distortion tensor one has \cite{iosifidis2020cosmological}
\begin{gather}
N_{\alpha\mu\nu}^{(n)}=X(t)u_{\alpha}h_{\mu\nu}+Y(t)u_{\mu}h_{\alpha\nu}+Z(t)u_{\nu}h_{\alpha\mu}\nonumber \\
+V(t)u_{\alpha}u_{\mu}u_{\nu} +\epsilon_{\alpha\mu\nu\lambda}u^{\lambda}W(t)\delta_{n,4}
\end{gather}	
where the functions $X(t),Y(t),Z(t),V(t),W(t)$ are linearly related to $\Phi(t),P(t),A(t),B(t),C(t)$ as it is trivially seen by employing 	($\ref{N1}$) (see \cite{iosifidis2020cosmological} for more details). These non-Riemannian degrees of freedom have a direct association with the microstructure of matter, once a hypermomentum tensor is given \cite{iosifidis2020cosmological} .We have now introduced the basic ingredients needed for the proof of our Theorem which we present below.

	\section{The non-Riemannian Raychaudhuri (or Acceleration) Equation}
The main result of this letter is contained in the following Theorem\footnote{In order to keep full generality and also demonstrate that there are no peculiarities arising for certain dimensions we will prove our result for general dimension $n$. Of course, in order to study physical Cosmology we have to set $n=4$ to our general result.  }.
\begin{theorem}
Consider an n-dimensional non-Riemannian FLRW spacetime. Then, the most general form of the acceleration equation  (aka Raychaudhuri eqn) in this spacetime reads
	\begin{gather}
\frac{\ddot{a}}{a}=-\frac{1}{(n-1)}R_{\mu\nu}u^{\mu}u^{\nu}+2\left( \frac{\dot{a}}{a} \right)\Phi +2\dot{\Phi} \nonumber \\
+\left( \frac{\dot{a}}{a} \right)\left(A+\frac{C}{2}\right)+\frac{\dot{A}}{2}-\frac{A^{2}}{4}-\frac{1}{4}AC  \nonumber \\
-A\Phi-C \Phi\label{Theorem}
\end{gather}
where $\Phi,A,B,C$ are the functions monitoring spacetime torsion/non-metricity as they appear in $(\ref{Scosm})$ and $(\ref{Qcosm})$. This is the most general form of the acceleration equation  when non-Riemannian effects (i.e. torsion and non-metricity) are also taken into account.
\begin{proof}
We start by the most general form of the Raychaudhuri equation in spaces with torsion and non-metricity\footnote{This is the most general non-Riemannian form of the Raychaudhuri equation. The case of vanishing non-metricity (i.e. torsion only) has been derived by several authors \cite{pasmatsiou2017kinematics,luz2017raychaudhuri,fennelly1991including}. }, as presented in \cite{iosifidis2018raychaudhuri,iosifidis2019metric}
\begin{gather}
\dot{\Theta}+\frac{\left(\Theta +\frac{a\cdot u}{l^{2}} \right)^{2}}{n-1}+\sigma^{2}-\omega^{2} -g^{\mu\nu}\nabla_{\mu}a_{\nu} \nonumber \\
-\frac{(a\cdot u)^{2}}{l^{4}}-2\frac{(a\cdot \xi)}{l^{2}}-u_{\alpha}Q^{\alpha\beta\mu}\nabla_{\beta}u_{\mu}+u_{\alpha}Q^{\mu\nu\alpha}\nabla_{\nu}u_{\mu}= \nonumber \\
=-R_{\mu\nu}u^{\mu}u^{\nu}+2 u_{\alpha}S^{\alpha\mu\nu}\nabla_{\nu}u_{\mu} \nonumber \\
+ 2 u^{\mu}u^{\beta}  \Big( g^{\nu\alpha} \nabla_{[\alpha}Q_{\beta]\mu\nu}-S_{\alpha\beta}^{\;\;\;\;\lambda}Q_{\lambda\mu}^{\;\;\;\;\alpha} \Big) \label{T}
\end{gather}
where $\Theta:=g^{\mu\nu}\nabla_{\mu}u_{\nu}$ is the expansion scalar, $\sigma_{\mu\nu}$ and $\omega_{\mu\nu}$ the shear and vorticity respectively. In addition $l$ is the length of the vector $u^{\nu}$ and $a_{\mu}:=u^{\lambda}\nabla_{\lambda}u_{\mu}$ represents the so-called hyperacceleration\cite{iosifidis2018raychaudhuri} , $\xi_{\mu}:=u^{\lambda}\nabla_{\mu}u_{\lambda}$ and $(a\cdot b):=a_{\mu}b^{\mu}$ is the dot product. Now, to begin with, we first consider a normalized $n$-velocity field $u_{\mu}u^{\nu}=-1$, that is $l=-1$\footnote{Notice that even for normalized velocity fields we still have a non-vanishing hyperacceleration which is not orthogonal to $u^{\mu}$, namely $u^{\mu}a_{\mu}\neq 0$}. Next we consider an FLRW background which implies that shear and vorticity vanish immediately ($\sigma_{\mu\nu}=0$, $\omega_{\mu\nu}=0$)  and the covariant forms of torsion and non-metricity are those in $(\ref{Scosm})$ and $(\ref{Qcosm})$.  Then  (\ref{expan}) becomes
\begin{gather}
\Theta=(n-1)\frac{\dot{a}}{a}+(n-1)X-V= \nonumber \\
=(n-1)\frac{\dot{a}}{a}+(n-1)\left( \frac{B}{2}- 2 \Phi -\frac{A}{2}\right)-\frac{C}{2}
\end{gather}
where we have also used that $\widetilde{\Theta}=\widetilde{\nabla}_{\mu}u^{\mu}=(n-1)\dot{a}/a$. It also holds that $\dot{\widetilde{\Theta}}=(n-1)\frac{\ddot{a}}{a}-(n-1)\frac{\dot{a}^{2}}{a^{2}}$, implying that
	\beq
\dot{\tilde{\Theta}}+\frac{\tilde{\Theta}^{2}}{n-1}=(n-1)\frac{\ddot{a}}{a}
\eeq
 which is the standard Riemannian part, while the rest of the terms will be of non-Riemannian nature. Using all the above and computing each term appearing in $(\ref{T})$ by using the aforementioned Cosmologicaly compatible forms of torsion and non-metricity   we arrive at the desired result (see Appendix for details) 
\begin{gather}
\frac{\ddot{a}}{a}=-\frac{1}{(n-1)}R_{\mu\nu}u^{\mu}u^{\nu}+2\left( \frac{\dot{a}}{a} \right)\Phi +2\dot{\Phi} \nonumber \\
+\left( \frac{\dot{a}}{a} \right)\left(A+\frac{C}{2}\right) +\frac{\dot{A}}{2}-\frac{A^{2}}{4}-\frac{1}{4}AC \nonumber \\
-A\Phi-C \Phi
\end{gather}

\end{proof}	
\textbf{Comment $1$.} The latter acceleration equation\footnote{Strictly speaking, at this point it is not an equation yet but rather an identity relating the various geometrical quantities of the space. It only becomes an equation once an action is fixed and the field equations are given relating the geometrical objects to their sources (matter). } is kinematic, namely its form (i.e. the functional dependence) is always the same regardless of the action one considers.

 \textbf{Comment $2$.} Note that one of the non-metricity functions $B(t)$ and the axial part of torsion $\zeta(t)$ are totally absent from the above  and therefore have no direct effect on the acceleration/deceleration of the Universe. They do, however, couple to the rest of the variables for a given Theory and therefore can indirectly affect the expansion rate.

 \textbf{Comment $3$.} Observe that on the right hand side of the acceleration equation appear terms proportional to $\dot{a}$ with non-Riemannian functions ($\Phi,A,C$) as their  coefficients. Then, 
 taking the mechanics analogue too far one could say the non-Riemannian effects produce some type of  'drag forces' which however could just as well speed up or slow down the expansion. 
  
\end{theorem}

Some additional remarks are the following. First	notice that the exact effect of torsion and non-metricity, when all the degrees of freedom are excited, is not quite clear and depends on various combinations. In the first line of the right-hand side of $(\ref{Theorem})$ we have the effect of torsion added to the standard $R_{\mu\nu}u^{\mu}u^{\nu}$ (which is the only one that appears in a Riemannian Geometry). In the second we separated out the purely non-metric contributions, and in the last line we have the combined effect of torsion and non-metricity appearing as couplings between their variables.

 We also observe that our general result $(\ref{Theorem})$ is in perfect agreement with the 2nd Friedmann equation with only torsion as appeared in \cite{kranas2019friedmann} and also with the general case presented in \cite{iosifidis2020cosmological}. In both of these studies the acceleration equation was obtained from the field equations of the Theory. Here we derived it kinematically (without referring to any action) and therefore this justifies the validity of our result. Given that these non-Riemannian degrees of freedom are dominant in the early Universe, they provide alternative mechanisms to cosmic Inflation\footnote{For a recent study on Inflaton fields in Metric-Affine Gravity, see \cite{shimada2019metric}}. In particular, from our derived acceleration equation and using the fact that $q=-\ddot{a}a/\dot{a}^{2}$ we readily find the deceleration parameter q in the presence of torsion and non-metricity
\begin{gather}
q=\frac{1}{(n-1)H^{2}}R_{\mu\nu}u^{\mu}u^{\nu}-2\frac{\Phi}{H} -2\frac{\dot{\Phi}}{H^{2}} \nonumber \\
-\frac{1}{H}\left(A+\frac{C}{2}\right)-\frac{\dot{A}}{2 H^{2}}+\frac{A^{2}}{4 H^{2}}+\frac{AC}{4 H^{2}}  \nonumber \\
+ \frac{A\Phi}{H^{2}}+ \frac{C \Phi}{H^{2}}  \label{q}
\end{gather}
 where as usual $H:=\dot{a}/a$ denotes the Hubble parameter.  The above equation expresses the behaviour of the deceleration parameter in the presence of non-Riemannian degrees of freedom. 

 As a final note before discussing applications, let us point out that the extra terms appearing on the right-hand side of would modify the singularity Theorem of Cosmology. It would then be very interesting to derive the version  of singularity Theorem that applies to ($\ref{Theorem}$). This would put constraints both on the Perfect Fluid characteristics of the hyperfluid ($\rho$ and $p$) and to its hypermomentum currents.

	\section{Applications}
	
	We shall now discuss the cases of pure torsion and pure non-metricity separately and also their combined effect in order to illustrate their significance on  cosmic acceleration.
	
	\subsection{Einstein-Cartan Geometry}
	For an Einstein-Cartan Geometry the non-metricity vanishes by construction and the acceleration equation reads
	\beq
	\frac{\ddot{a}}{a}=-\frac{1}{(n-1)}R_{\mu\nu}u^{\mu}u^{\nu}+2\left( \frac{\dot{a}}{a} \right)\Phi +2\dot{\Phi} \label{t}
	\eeq
	which is in perfect agreement with \cite{kranas2019friedmann}\footnote{Note a sign difference between our definitions of $\Phi$ here with the $\phi$ of \cite{kranas2019friedmann}.}. We see that in contrast to non-metricity, torsion changes the acceleration in a minimal way. Defining $f(t):=a(t)\Phi(t)$ we observe that torsion aids to acceleration when $f(t)$ is strictly increasing ($\dot{f}>0$), slows down expansion when $\dot{f}<0$ and has no effect on the acceleration whenever $f=const.$ as is readily seen from the above. Of course, when constructing singularity Theorems, the whole contribution (including $-R_{\mu\nu}u^{\mu}u^{\nu}$) from the right-hand side of $(\ref{t})$ must taken into account. We see however that the possibility of singularity avoidance could occur through the help of torsion (see  also \cite{trautman1973spin} in relation to this point).

	\subsection{Symmetric Geometry (Vanishing Torsion)}
	Let us now consider the case of vanishing torsion and keep only non-metricity. Then, our acceleration equation takes the form
	\begin{gather}
	\frac{\ddot{a}}{a}=-\frac{1}{(n-1)}R_{\mu\nu}u^{\mu}u^{\nu}+\left( \frac{\dot{a}}{a} \right)\left( A+\frac{C}{2} \right) \nonumber \\
	+\frac{\dot{A}}{2}-\frac{A^{2}}{4} -\frac{AC}{4}
	\end{gather}
	We see that now the situation is more complicated from the previous pure torsion case and no immediate conclusion can be drawn since there are two functions evolved. The only term with a fixed sign is $-A^{2}/2$ which always decelerates the expansion. However, concrete results about the net effect on the expansion can only be obtained when one knows the exact form of $A$ and $C$, which of course depend on the Theory considered.
	Interestingly, when non-metricity is restricted to be of the Weyl type ($A=-C$,$B=0$), the above equation reduces to 
		\beq
	\frac{\ddot{a}}{a}=-\frac{1}{(n-1)}R_{\mu\nu}u^{\mu}u^{\nu}+\left( \frac{\dot{a}}{a} \right)\frac{A}{2}+\frac{\dot{A}}{2} \label{t1}
	\eeq
	Note that the latter is identical to $(\ref{t})$ upon the exchange
		\beq
	2 \Phi \leftrightarrow \frac{A}{2} \label{dual}
	\eeq
	This is the torsion/non-metricity duality that has been reported earlier in the literature \cite{iosifidis2019torsion} (see also \cite{iosifidis2018raychaudhuri,klemm2020einstein}) which holds true between vectorial torsion and Weyl non-metricity. Therefore, in this case the  effect on the acceleration is identical to the pure torsion case.

	\subsection{Weyl-Cartan Geometry}
	For a Weyl-Cartan Geometry we have arbitrary torsion but the non-metricity is of the Weyl type, that is $Q_{\alpha\mu\nu}=\frac{Q_{\alpha}}{n}g_{\mu\nu}$. In this case ($\ref{Theorem}$) becomes
	\begin{gather}
	\frac{\ddot{a}}{a}=-\frac{1}{(n-1)}R_{\mu\nu}u^{\mu}u^{\nu} \nonumber \\+\left( \frac{\dot{a}}{a} \right)\left( 2\Phi+\frac{A}{2}\right) +2\dot{\Phi}+\frac{\dot{A}}{2} \label{t2}
	\end{gather}
	Note that this is invariant upon the duality exchange ($\ref{dual}$). Again, the right hand side has the same form with the pure torsion case and the pure Weyl-non-metricity but now for the 'Affine' function $Y=2\Phi+A/2$  in place of $\Phi$ and $A$ alone.

	\section{Conclusions}
	 We have derived the most general form of the non-Riemannian  cosmic acceleration equation (Raychaudhuri) in FLRW Universes. We have therefore provided a helpful tool allowing one to decode the effects of torsion and non-metricity in an expanding homogeneous Universe. In the general case when all non-Riemannian degrees of freedom are excited, there are no fixed sign terms  and therefore it is not clear what would be the impact of  torsion and and non-metricity separately and also their net effect. Concrete results can only be obtained once an action is given. Interestingly, in contrast to the classical Perfect Fluid case where the combination $\rho+3p$ has a definite positive sign for standard forms of matter, and a fixed negative sign for dark energy, here the non-Riemannian contributions could experience a change in sign. This would mean that they could aid to acceleration at a given period but slow it down at another stage (and vice versa of course). These are all possibilities and concrete conclusions may only be drawn once we adopt to a given Theory.

	 It is important to note the fact that  the acceleration (Raychaudhuri) equation $(\ref{Theorem})$ is kinematic (i.e. its form is the same regardless the action) means that it is completely generic and holds for any Metric-Affine Theory given that the background geometry is FLRW. This opens up the possibility to systemically study the Cosmology of the various Metric-Affine Theories and rule out the ones that are in disagreement with observations.
	 Therefore, having obtained equation ($\ref{Theorem}$) the next logical step would be to consider a specific Metric-Affine Theory, in the presence of a Cosmological hyperfluid \cite{iosifidis2019torsion}.\footnote{Other hyperfluid models include \cite{obukhov1993hyperfluid,obukhov1996model,babourova1998perfect}. Note that we can also have geometrically induced torsion and non-metricity without the presence of a hyperfluid. This however would require to adopt non-standard Gravitation Lagrangians.} This is currently under investigation.

	\section{Acknowledgments}
	I would like to thank Christos Tsagas for providing some useful references and some discussion.
	This research is co-financed by Greece and the European Union (European Social Fund- ESF) through the
	Operational Programme 'Human Resources Development, Education and Lifelong Learning' in the context
	of the project “Reinforcement of Postdoctoral Researchers - 2
	nd Cycle” (MIS-5033021), implemented by the
	State Scholarships Foundation (IKY).

	\appendix
	
\section{Supplementary material for the proof}
	Using the covariant forms ($\ref{Scosm}$) and $(\ref{Qcosm})$, considering a co-moving observer (i.e. $u^{\mu}=\delta^{\mu}_{0}$) and keeping in mind that $\dot{}=u^{\mu}\nabla_{\mu}$, we have
		\beq
	a_{\mu}=\dot{u}_{\mu}=-\frac{C}{2}u_{\mu}\;\;, \;\; 	A^{\mu}=\dot{u}^{\mu}=\frac{C}{2}u^{\mu}\;\;, 	
	\eeq
	\beq
	a_{\mu}u^{\mu}=\xi_{\mu}u^{\mu}
	\eeq
	\beq
	(a\cdot \xi)=-\frac{C^{2}}{4}\;\;,\;\; 	
	a_{\mu}u^{\mu}=-u_{\mu}A^{\mu}
	\eeq
		\beq
	a_{\mu}u^{\mu}=-\frac{1}{2}Q_{\alpha\mu\nu}u^{\alpha}u^{\mu}u^{\nu}=V=\frac{C}{2}
	\eeq
		\beq
	g^{\mu\nu}\nabla_{\mu}a_{\nu}=-\left( \frac{1}{2}\dot{C}+\frac{C}{2}\Theta \right)
	\eeq
	where
	\beq
A^{\mu}:=u^{\alpha}\nabla_{\alpha}u^{\mu}	
	\eeq
	\beq
	a_{\mu}:=u^{\alpha}\nabla_{\alpha}u_{\mu}
	\eeq
	\beq
	\xi_{\mu}:=u^{\alpha}\nabla_{\mu}u^{\alpha}
	\eeq
	Note also the relation between the two accelerations
		\beq
	a_{\mu}=A_{\mu}-Q_{\alpha\mu\nu}u^{\alpha}u^{\nu}
	\eeq
	The above are derived by setting $l=1$ at the identities appearing in\cite{iosifidis2018raychaudhuri}. Continuing we compute
		\beq
	S_{\alpha}=(n-1)\Phi u_{\alpha}
	\eeq
		\beq
	Q_{\alpha}=\Big[ (n-1)A -C \Big]u_{\alpha}
	\eeq
	\beq
	S_{\alpha\mu\beta}u^{\mu}=\Phi h_{\alpha\beta}
	\eeq
		\beq
	Q_{\alpha\mu\nu}	u^{\nu}=-\left[ \frac{B}{2}h_{\alpha\mu}+Cu_{\alpha}u_{\mu} \right]
	\eeq
	\beq
	\tilde{Q}_{\alpha}=\Big[ \frac{(n-1)}{2}B -C \Big] u_{\alpha}
	\eeq
		\beq
	u_{\alpha}Q^{\alpha\mu\nu}\nabla_{\mu}u_{\nu}=-A\Theta-\frac{C}{2}(A+C)
	\eeq
	\beq
	u_{\alpha}Q^{\mu\nu\alpha}\nabla_{\nu}u_{\mu}=-\frac{B}{2}\Theta-\frac{C}{2}\Big( \frac{B}{2}+C \Big)
	\eeq
	\beq
	u_{\alpha}S^{\alpha\mu\nu}	\nabla_{\nu}u_{\mu}=-\Phi \Theta-\frac{C}{2}\Phi 
	\eeq
		\begin{gather}
	u^{\mu}u^{\beta}g^{\nu\alpha}\nabla_{\alpha}Q_{\beta\mu\nu}=	\Big( A+\frac{B}{2}+C \Big)  \Theta \nonumber \\
	 +u_{\mu}\dot{u}^{\mu}\Big( A+\frac{B}{2}+2 C \Big)-C\tilde{Q}_{\mu}u^{\mu}+\dot{C}
	\end{gather}
	\begin{gather}
	u^{\mu}u^{\beta}g^{\nu\alpha}\nabla_{\beta}Q_{\alpha\mu\nu}=-	\frac{(n-1)}{2}\dot{B}+\dot{C} \nonumber \\
	-\frac{(n-1)}{2}AB +\frac{(n-1)}{4}BC -\frac{3}{2}C^{2}
	\end{gather}
	
	Plugging all of these into the Raychaudhuri equation after some rather lengthy but straightforward calculations we arrive at $(\ref{Theorem}) $. Note that in the final result, the $\dot{B}$ terms have canceled out.

	\bibliographystyle{unsrt}
	\bibliography{ref}

		\end{document}